# Using a Big Data Database to Identify Pathogens in Protein Data Space

Ashley Conard[1,2,3], Stephanie Dodson[1,2,4], Jeremy Kepner[1,2,5,6], Darrell Ricke[1,2]
[1]MIT Lincoln Laboratory, Lexington, Massachusetts; [2]MIT BeaverWorks Center, Cambridge, Massachusetts;
[3]Université Libre de Bruxelles, Brussels, Belgium; [4]Brown University, Providence, Rhode Island;
[5]MIT Mathematics Dept., Cambridge, Massachusetts; [6]MIT CSAIL, Cambridge, Massachusetts

*Abstract*—Current metagenomic analysis algorithms require significant computing resources, can report excessive false positives (type I errors), may miss organisms (type II errors / false negatives), or scale poorly on large datasets. This paper explores using big data database technologies to characterize very large metagenomic DNA sequences in protein space, with the ultimate goal of rapid pathogen identification in patient samples. Our approach uses the abilities of a big data databases to hold large sparse associative array representations of genetic data to extract statistical patterns about the data that can be used in a variety of ways to improve identification algorithms.

Keywords-Bioinformatics; Big Data; Accumulo; Databases

## I. Introduction

The current Ebola outbreak, which originated in Guinea, has now become a worldwide concern [Ebola 2015]. Physicians and researchers are unable to identify the virus until certain symptoms become apparent, and many people in Africa have mistaken Ebola for other diseases such as malaria. Health workers, family members, and mourners are all equally at risk. Early detection and identification are essential for patients and for those in contact with infected individuals. To address this type of problem, big data database technologies [Change et al 2008] such as SciDB [Balazinska et al 2009] and Apache Accumulo [Wall, Cordova & Rinaldi 2013] can assist with protein identification for characterizing the genes of pathogens. Protein analysis is useful to identify toxin genes, virulence genes, antibiotic resistance genes, and pathogenicity genes.

The goal of this research project is to quickly diagnose a patient with any kind of infection and inform him or her in just a few hours. Currently, technology exists to quickly and accurately perform organism identification on few DNA segments. Unfortunately, the time to actionable data increases significantly with an increase in segment number. In a clinical setting, it can take from 6 to 14 days to acquire actionable data from the time of an event [Koch 2011]. We present a protein-based approach related to our prior work [Kepner, Ricke & Hutchison 2013; Conard et al 2013; Dodson, Ricke & Kepner 2014]. This new protein identification algorithm relies on the ability of the Accumulo database to store large sparse associative array representations of genetic data and to extract statistical patterns about the data that can be used in a variety of ways to improve algorithms. Results are shown from various *in silico* datasets, and compared with results from Basic Local Alignment Search Tool (BLAST) and the U.S. Defense Threat Reduction Agency (DTRA) "Identify Organisms from a Stream of DNA Sequences" Challenge [Rosenzweig et al 2015].

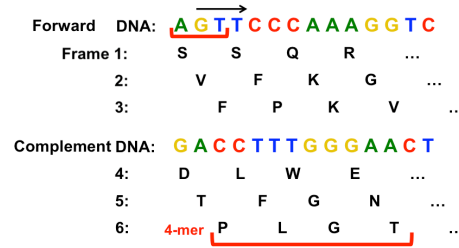

Figure 1. Sliding window method identifies all possible proteins for the forward and complement DNA strands. Each protein sequence is parsed into 4-mer words (e.g. PLGT) and used to match against protein reference sequences. Transcription step not required.

## II. Approach

A genetic sample from an organism consists of many distinct strings of DNA that are coded in four letters (A, C, G, and T), corresponding to DNA's four bases. A typical sample may consist of thousands or millions of strings containing hundreds or thousands of bases. Three DNA bases map onto one of 20 amino acids in protein space. Thus, DNA can be mapped onto proteins by looking at each of the six possible translation frames. Amino acids are grouped into k-mers with a "sliding window" procedure (see Figure 1) to create distinct signatures in a vector with the k-mer as the index. Figure 1 shows grouping of the data in 4-mers consisting of four amino acids. The total number of possible 4-mers is $20^4 = 160,000$, of which some are more or less common in each segment.

Each sequence of DNA can thus be mapped into a sparse associative array whereby the rows are the sequence label and columns consist of each 4-mer word. Big data databases such as Apache Accumulo are ideally suited for storing large sparse associative arrays [Kepner et al 2013] of genetic data and allow for rapid row or column look-ups. Appache Accumulo has sufficient ingest performance [Kepner et al 2014] to keep up with the large volumes of data generated by genetic sequence machines. Moreover, by using an associative array computation technology such as the Dynamic Distributed Dimentional Data Model (D4M) (d4m.mit.edu)[Kepner et al 2014], it is possible to compare a large collection of genetic data via fast and simple sparse matrix products, usually an $N^2$ operation.

An additional feature of Apache Accumulo is its built-in ability to accumulate sums while data is being ingested. Accumulation allows the statistical distribution of the rows and columns to be used in subsequent analysis. In the case of the protein amino acids, the frequency of different amino acid 4-mers in specific populations can be a useful tool for enhancing the speed and accuracy of different identification algorithms.

### III. RESULTS & CONCLUSIONS

The distribution of 4-mer codons for selected bacteria and viruses from known reference databases as computed using D4M and Accumulo are shown in Figures 2 and 3. These distributions are a natural byproduct of using the standard schemas for ingesting data into Accumulo [Kepner et al 2013].

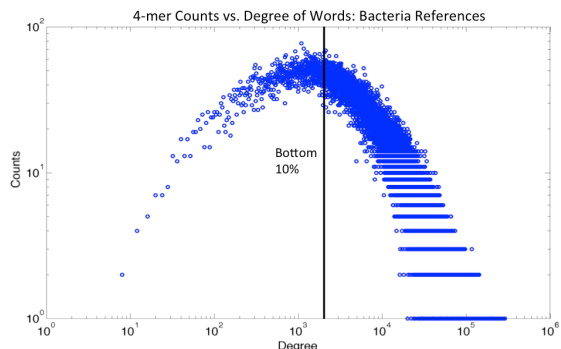

Figure 2. Degree distribution of 4-mers for bacteria reference database. Vertical line shows threshold for bottom 10% of all data.

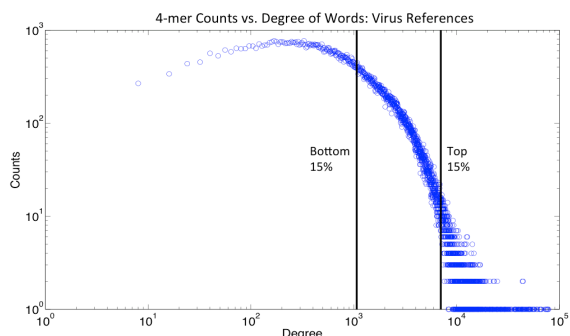

Figure 3. Degree distribution of 4-mers for virus reference database. Vertical lines show thresholds for bottom 15%, and top 15% of data.

In random sequences, all words occur in the same frequency. Frequently used words are considered to have a high degree and are present in many organisms. This introduces false positives, and reduces algorithm performance. Protein sequences contain common and uncommon words based on necessary differences that influence structure and function in the body. Words that have a high degree are supernodes. Removing these supernodes leaves the more rarely used words, reducing protein identification search space.

Sub-sampling techniques that removed various 4-mers based on their frequency were tested on two datasets. Dataset 1 was generated from a human sample spiked with two strains of the *in silico* bacteria organism Escherichia coli (E. coli), using FastqSim [Scherbina 2014]. Dataset 2 was generated in the same fashion as Dataset 1, but was spiked with 22 viruses in varying quantities. The reference sets were created from DNA found in GenBank. Figures 2 and 3 show thresholds (chosen to explore space) used to subsample. Both sets of sub-sampling results were examined for organism identification accuracy.

For Dataset 1, an extensive algorithm (without subsampling) was first implemented, and results were compared with BLAST findings at ≥70% sequence identity, and with DTRA algorithms. We correctly identified 97.9% of the E. coli samples, where BLAST detected 78%. The extensive algorithm was also comparable with six of the top DTRA algorithms, of which Metaphyler did not detect E.coli and had a high run time [Rosenzweig et al 2015]. Using accelerated sub-sampling, even with only the bottom 10% of the data, our algorithm was able to detect the majority of the reads correctly, with few false positives. In Dataset 2, the degree distribution in the virus reference dataset shows that the most common 4-mers can also help discern and classify sequences because viruses are mutable; thus, these supernodes vary. Our extensive algorithm correctly detected 22 viruses with few false positives. When sub-sampling, we observed that the bottom 15% and the top 15% are equally useful and quick to identify organisms, with the bottom 15% taking slightly less time. Both sub-sampling results correctly identified 22 viruses with few false positives.

For both bacteria and viruses, the degree information provided by the Accumulo database allows the most valuable genetic reference data to be selected while preserving the usefulness of the data in organism identification. The time to execute the identification algorithms is proportional to the amount of data that needs to be compared and so a corresponding speedup in the computation was also seen.